\documentclass[12pt]{article}

\usepackage{epsfig}

\begin{document}

\begin{center}
{\bf\large Incommensurate Spin--Peierls Phases\\ 
           in the One--Dimensional Quantum Isotropic $XY$ Model}

OLEG DERZHKO, TARAS KROKHMALSKII\\
Institute for Condensed Matter Physics\\
 1 Svientsitskii Street, L'viv--11, 79011, Ukraine
\end{center}

\abstract{
Using the numerical approach 
for a study of the thermodynamic properties 
of the nonuniform one--dimensional spin-$\frac{1}{2}$ isotropic $XY$ model 
in a transverse field 
we examine different lattice distortions
to reveal which spin--Peierls phases are realized in the magnetic chain
at zero temperature in the presence of external field.}

\vspace{10mm}

An interest in the theoretical study of quantum spin chains 
exhibiting spin--Peierls phases 
has incredibly grown  
since the discovery of the first inorganic spin--Peierls compound CuGeO$_3$
(for a review see$^1$).
Although the quantum Heisenberg model 
is usually used as an appropriate model 
to describe the spin--Peierls phase transition in the available materials  
some generic features can be clarified   
within the framework of the simpler spin-$\frac{1}{2}$ isotropic $XY$ chain
(see$^{2,3}$ and references therein).
The latter spin model can be reformulated 
using the Jordan--Wigner transformation 
as a one--dimensional model of tight--binding spinless fermions.
As a result, 
the exhaustive analytical and numerical analysis of different properties of the model 
becomes possible.

In what follows 
we analyze a stability of various spin--Peierls phases at zero temperature
in the presence of external field.
For this purpose we consider 
a nonuniform spin-$\frac{1}{2}$ isotropic $XY$ chain in a transverse field 
defined by the Hamiltonian
\begin{eqnarray}
\label{001}
H=\sum_{n=1}^{N-1}
J_n\left(s_n^xs_{n+1}^x+s_n^ys_{n+1}^y\right)
+\Omega\sum_{n=1}^{N}s_n^z.
\end{eqnarray}
Here $J_n=J\left(1+\delta_n\right)$ 
is the nonuniform exchange interaction between the sites $n$ and $n+1$
and the sequence of parameters
$\delta_1,\ldots,\delta_{N-1}\equiv\left\{\delta_n\right\}$ 
defines a certain lattice distortion. 
$\Omega$ is the value of a uniform external field directed along $z$ axis. 
Using the numerical approach described in detail in$^4$
we calculate the ground state energy 
$E_0(\left\{\delta_n\right\},\Omega)=Ne_0(\left\{\delta_n\right\},\Omega)$
of the spin chain (\ref{001}).
The ground state energy does not depend on the sign of $J$;
we fix the units putting in what follows $\vert J\vert =1$.
We accurately analyze the finite--size effects 
to be sure that our results pertain to the thermodynamic systems.
Taking $N=1000$ we no more observe the finite--size effects 
which are nicely pronounced when $N=100$ or less.

In the adiabatic treatment of the spin--Peierls instability at zero temperature 
one should examine the total energy 
which consists of the magnetic part $E_0(\left\{\delta_n\right\},\Omega)$
and the elastic part $\alpha\sum_{n=1}^{N}\delta_n^2$ 
for different lattice configurations $\left\{\delta_n\right\}$.
Here $\alpha$ is the parameter which measures the lattice stiffness.
The total energy per site 
will be denoted by ${\cal{E}}(\left\{\delta_n\right\},\Omega)$.

To begin with, 
we assume $\alpha=0.5$
and examine the dimerization ansatz 
$\delta_n=-\delta(-1)^n$,
$0\le\delta\le 1$.
${\cal{E}}(\delta,\Omega)-{\cal{E}}(0,\Omega)$ 
is shown in Fig. 1a.
\begin{figure}
\begin{center}
\epsfclipon
\hspace*{0pt}
\epsfig{file=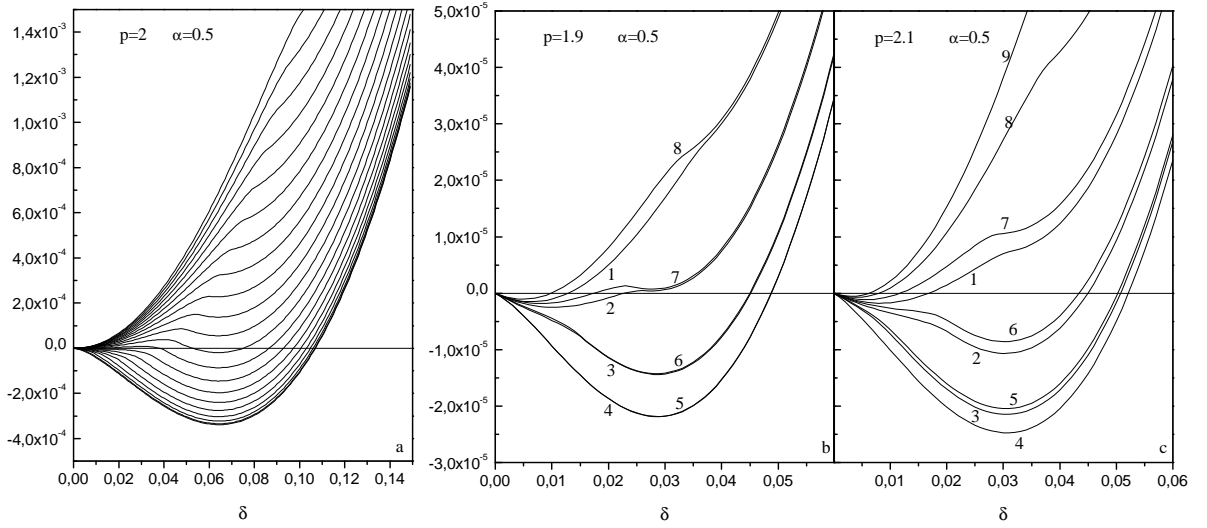,width=16cm}
\parbox{16.cm}
{\caption{
\small
{${\cal{E}}(\delta,p,\Omega)-{\cal{E}}(0,p,\Omega)$ vs. $\delta$ at different $\Omega$,
$\alpha=0.5$. 
(a) $p=2$, $\Omega=0.000,\;0.005,\;\ldots,\;0.100$ (from bottom to top);
(b) $p=1.9$, $\Omega=0.065$ (1), $0.070$ (2), $\ldots,$ $0.100$ (8);
(c) $p=2.1$, $\Omega=0.060$ (1), $0.065$ (2), $\ldots,$ $0.100$ (9).}}}
\end{center}
\end{figure}
Inspecting the displayed curves one concludes that  
i)
if $\Omega$ does not exceed $\Omega_a$ ($\approx 0.035$) 
only the dimerized phase occurs
(with the dimerization parameter $\delta^{\star}\approx 0.065$);
ii) 
if $\Omega$ exceeds $\Omega_a$ but does not exceed $\Omega_b$ ($\approx 0.045$) 
both the dimerized and uniform phases are possible,
however, the former phase is favorable;
iii) if $\Omega$ exceeds $\Omega_b$ but does not exceed $\Omega_c$ ($\approx 0.065$) 
both the dimerized and uniform phases are possible,
however, the latter phase is favorable;
iv) if  $\Omega$ exceeds $\Omega_c$ 
only the uniform phase is possible.
Such a behavior of the total energy corresponds to a scenario 
of the first order phase transition driven by the external field.
Within the frames of the adopted ansatz for $\delta_n$
we are restricted to the dimerized and uniform phases 
and cannot judge about a possibility  
of more complicated lattice distortions.

To discuss 
whether the total energy can be lowered by another 
(not uniform) lattice pattern as the field increases
we introduce a trial distortion of the form
\begin{eqnarray}
\label{002}
\delta_n=-\delta\cos\left(\frac{2\pi}{p}n\right),
\end{eqnarray}
where $p$ is the period of modulation 
(e.g., $p=2$ yields the dimerization ansatz).
A behavior of ${\cal{E}}(\delta,p,\Omega)$
for $p=1.9$ and $p=2.1$ as the field increases 
can be seen in Figs. 1b and  1c, respectively.
From the displayed plots one concludes that a long--period structure
does arise, if $\Omega$ exceeds the value about $0.06$.
Thus, the dimerized phase transforms into a long--period phase
rather than into the uniform phase while the field increases.
 
Further, let us clarify
whether for {\em any small field} there exists such $p$ 
which yields the total energy lower than that for the dimerized chain.
For a certain $\alpha$
at the fixed $\Omega$ ($=0,\;0.025,\;0.05,\;\ldots$)
we examine the dependence ${\cal{E}}(\delta,p,\Omega)$ vs. $\delta$
seeking the minimal value of ${\cal{E}}(\delta,p,\Omega)$ 
for different $p$ (in a sufficiently large region)
and then compare those minimal values. 
As a result, 
we find the value of $p$ 
of the most energetically favorable lattice distortion (\ref{002}).
It will be denoted as $p^{\star}$.
We repeat the search of the most favorable $p$ at the fixed $\Omega$ 
for different $\alpha$. 
The results obtained are collected in Fig. 2.
\begin{figure}
\begin{center}
\epsfclipon
\hspace*{0pt}
\epsfig{file=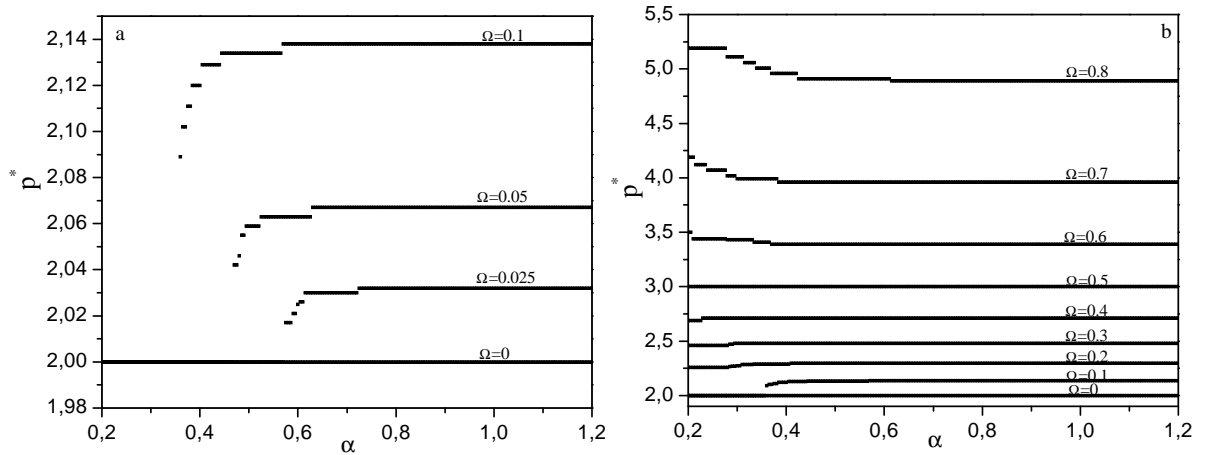,width=16cm}
\parbox{16.cm}
{\caption{
\small
{$p^{\star}$ vs. $\alpha$ for several values of $\Omega$.}}}
\end{center}
\end{figure}
From Fig. 2a one can learn,
for example,
that for $\alpha=0.5$ the dimerized phase is favorable at least up to $\Omega=0.025$
whereas at $\Omega=0.05$ it is already unstable 
with respect to the transformation into a long--period phase 
(with $p$ about $2.06$).
The important conclusion that can be drawn from Fig. 2a
is that the dimerized phase for any $\alpha$ 
{\em persists} up to a certain characteristic field 
(the value of which decreases as $\alpha$ increases).
Another evident observation 
is that at zero field 
the dimerized phase brings the lowest total energy for any $\alpha$
(that agrees with the result proved in$^5$).
 
With the increase of the field the lattice parameterized by (\ref{002})
may exhibit short--period phases,
for example, the trimerized phase$^6$ for which $p=3$.
However, as we shall see below, 
a behavior of the trimerized phase is essentially different 
in comparison with that of the dimerized phase.  
Fig. 3a demonstrates 
that the trimerized phase may 
really occur at the field about $0.5$.
\begin{figure}
\begin{center}
\epsfclipon
\hspace*{0pt}
\epsfig{file=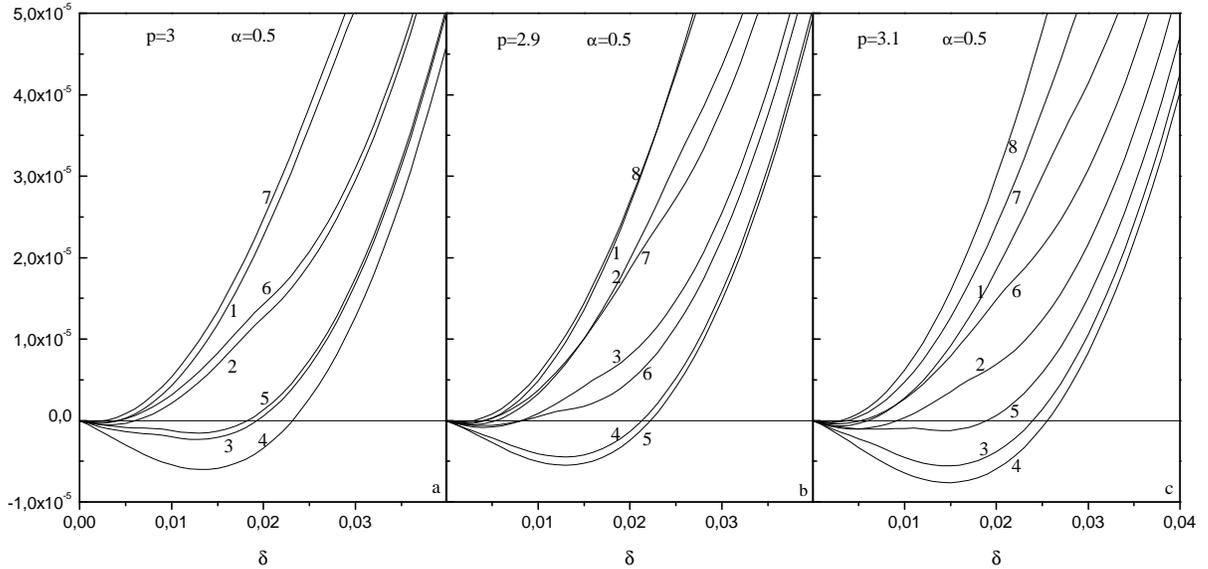,width=16cm}
\parbox{16.cm}
{\caption{
\small
{${\cal{E}}(\delta,p,\Omega)-{\cal{E}}(0,p,\Omega)$ vs. $\delta$ at $\Omega$ about 0.5,
$\alpha=0.5$. 
(a) $p=3$,   $\Omega=0.485$ (1), $0.490$ (2), $\ldots,$ $0.515$ (7);
(b) $p=2.9$, $\Omega=0.450$ (1), $0.455$ (2), $\ldots,$ $0.485$ (8);
(c) $p=3.1$, $\Omega=0.515$ (1), $0.520$ (2), $\ldots,$ $0.550$ (8).}}}
\end{center}
\end{figure}
In the same region of fields, 
various long--period structures are possible
(Figs.  3b,  3c).
Moreover, for any small deviation of the field from 0.5
there exists such $p$ 
for which the lattice distortion (\ref{002}) gives smaller energy than for $p=3$.
This can be seen in Fig. 2b. 
Thus, contrary to the dimerized phase,
the trimerized phase does not persist with the field varying 
and so it continuously transforms into a certain long--period phase.

It is worth noting
that using the exact analytical expression$^3$ for  
$e_0(\left\{\delta_n\right\},\Omega)$
for the spin-$\frac{1}{2}$ isotropic $XY$ chain in a transverse field of period 3
one can find that there may exist two lattice distortions,
which preserve the chain length ($\delta_1+\delta_2+\delta_3=0$), 
for which ${\cal{E}}(\left\{\delta_n\right\},\Omega)$ may have an extremum. 
Namely, $\delta_1=\delta_2$ (it follows from (\ref{002}) if $p=3$)  
and $\delta_1=-\delta_2$.
We check numerically that in both cases 
${\cal{E}}(\left\{\delta_n\right\},\Omega)$
really exhibits a minimum at $\Omega$ about 0.5
(see Fig. 4).
\begin{figure}[ht]
\begin{center}
\epsfclipon
\hspace*{0pt}
\epsfig{file=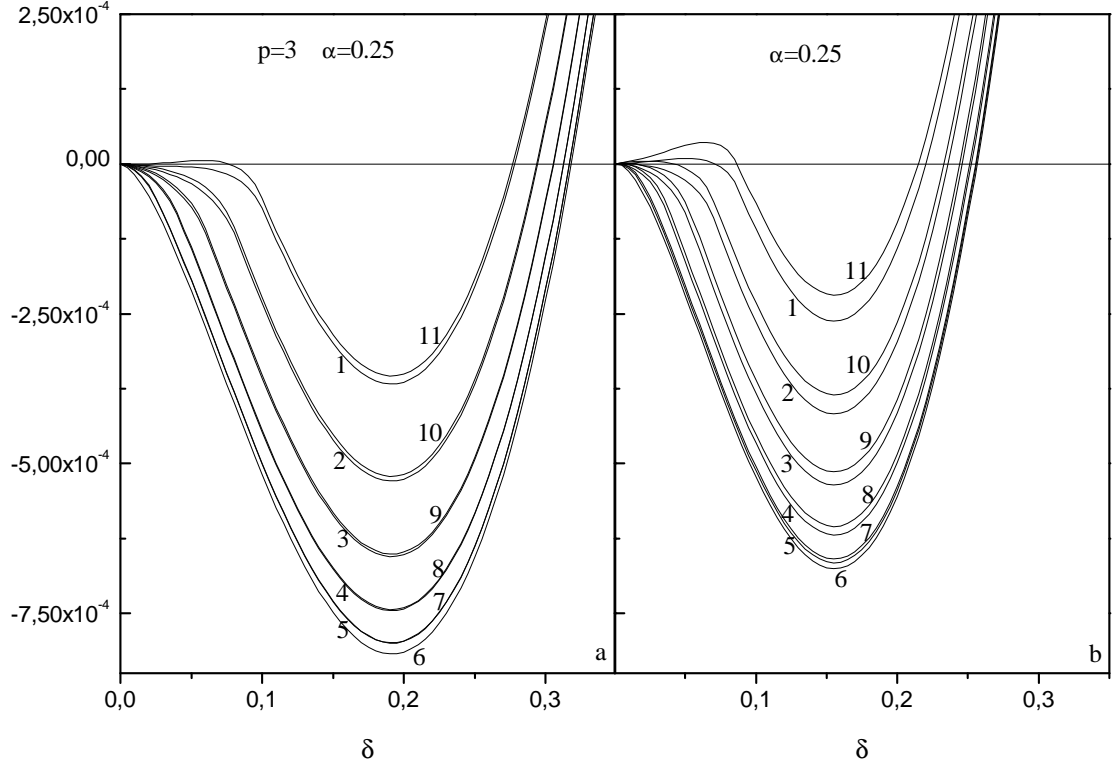,width=15cm}
\parbox{16.cm}
{\caption{
\small
{${\cal{E}}(\{\delta_n\},\Omega)-{\cal{E}}(\{0\},\Omega)$ 
vs. $\delta$ 
for two lattice distortions of period 3,
i.e.
$\delta_1=\delta_2=\frac{1}{2}\delta$,
$\delta_3=-\delta$ (a)
and
$\delta_1=-\delta_2=\delta$,
$\delta_3=0$ (b),
at $\Omega$ about 0.5,
$\alpha=0.25$. 
$\Omega=0.45$ (1), $0.46$ (2), $\ldots,$ $0.55$ (11).}}}
\end{center}
\end{figure}
However, due to the instability 
with respect to the transformation into long--period phases 
those trimerized patterns 
are not so important as the dimerized configuration.

To summarize,
using a numerical approach we examined (within the adiabatic approximation) 
a stability of various lattice distortions 
of the spin-$\frac{1}{2}$ isotropic $XY$ chain at zero temperature
in the presence of external field.
We found 
that while the field increases 
the favorable at zero field dimerized phase persists 
until the field achieves a certain characteristic value,
at which a first order transition into incommensurate phase occurs. 
Although in a moderate field one can find distortions having short periods 
(for example, of period 3) 
which bring a gain in the total energy,
any small variation of the field leads 
to more energetically favorable long--period structures.  
In strong fields, 
the uniform lattice can be expected.
Clearly,
since we are restricted to the adopted ansatz for a lattice distortion (\ref{002}) 
we can say for sure what lattice distortion is not realized
rather than to point out which lattice distortion should occur.
Finally, let us note that the elaborated numerical procedures 
can be easily applied to other distortion patterns,
e.g., an array of solitons$^7$.

\vspace{3mm}

The authors thank J.~Richter and O.~Zabyrannyi for discussions.
O. D. acknowledges the kind hospitality of the Magdeburg University 
in the summer of 2000
when a part of this work was carried out.

\vspace{5mm}

\noindent {\bf References}

\noindent 1. 
Boucher J. P., Regnault L. P.,
J. Phys. I France, {\bf 6,} 1939 (1996).

\noindent 2.  
Taylor J. H., M\"{u}ller G., 
Physica A, {\bf 130,} 1 (1985).

\noindent 3.
Derzhko O., Richter J., Zaburannyi O.,
Physica A, {\bf 282,} 495 (2000).

\noindent 4.
Derzhko O., Krokhmalskii T.,
phys. stat. sol. (b), {\bf 208,} 221 (1998).

\noindent 5.
Kennedy T., Lieb E. H.,
Phys. Rev. Lett., {\bf 59,} 1309 (1987).

\noindent 6.
Okamoto K.,
Solid State Commun., {\bf 83,} 1039 (1992).

\noindent 7.
Feiguin A. E., Riera J. A., Dobry A., Ceccatto H. A.,
Phys. Rev. B, {\bf 56,} 14607 (1997).

\end{document}